# Causal Evaluation of Contributing Factors to Urban Heat Island

**Sadaf Alavi Soltani, Saeid Niazmardi[1], Ali Esmaeily**

Department of Surveying Engineering, Faculty of Civil and Surveying Engineering,
Graduate University of Advanced Technology, Kerman, Iran

## Abstract

Urban heat islands (UHI) are formed due to complex interactions between various factors. UHI, its contributing factors, and their interaction vary over time and location. Accordingly, understanding the causal relation between UHI and its contributing factors is essential to minimizing its adverse effects on the environment and human health. Here, we proposed a statistical method based on Hotelling's T-square test to analyze this association. The proposed test estimates the UHI trends across different urban districts and compares the UHI contributing factors between the districts with increasing and non-increasing UHI trends. This comparison, if significantly different, can be interpreted as evidence of a causal association between the factor and UHI. This research used the proposed test to analyze the UHI and its contributing factors across 22 municipal districts of Tehran between 2003 and 2021. We examined the time series of weather conditions (measured by precipitation, NDSI, and NDWI), vegetation cover (measured by NDVI and EVI), and urban density (measured by NDBI) as factors contributing to the UHI, which was measured through nighttime LST. The results showed that all districts in Tehran exhibited stable or increasing trends in LST, leading to UHI effects. The proposed test indicated that the temporal changes in NDWI and NDBI did not have a causal relationship with UHIs. Meanwhile, variations in other factors were identified as contributing to the intensification of UHIs.

**Keywords**: Urban heat island, Casual analysis, Correlation analysis, Hotelling's T-square, UHI's contributing factors.

[1] Correspomding author (s.niazmardi@kgut.ac.ir)

## 1. Introduction

Urban Heat Island (UHI) occurs when temperatures in urban areas are higher than in the surrounding rural areas [1]. UHI poses major threats to urban environments and public health. Higher energy consumption [2, 3], increased water usage [4], exacerbated heat-related health issues [5], and elevated air pollution levels [6] are just a few examples of UHI's adverse effects. Various factors contribute to the formation or intensity of the UHIs over time. These factors can broadly be categorized as environmental, socio-economic, and political. Among these, the environmental factors such as urban density, urban green spaces area, and water bodies have the most direct effect on UHI. Radiated heat from fuel combustion and urban transportation systems [7], heat retention by urban surfaces [8], and trapped warm air beneath cold rooftops during winter or at night can also contribute to UHIs [9]. These factors have complex and time-varying interactions with the UHI. Understanding this interaction is crucial for minimizing UHI's effects. Remote sensing data, with their ability to continuously measure land surface temperature (LST), are among the most commonly used data that can be used to analyze the association between the UHI and its contributing factors.

Numerous studies have been published on using remote sensing data to analyze UHIs and their contributing factors. For instance, in [10, 11], the researchers used regression and correlation methods to identify the relationship between LST and precipitation data. Water bodies and snow are two contributing factors that may influence the formation of UHIs. Using the correlation analysis, the relationship between LST and Normalized Difference Water Index (NDWI) and Normalized Difference Snow Index (NDSI) was studied by [11-13].

The amount of urban green space is a crucial factor contributing to UHIs. Thus, the relation between several vegetation indices and LST has been analyzed using the correlation method.

Among the most commonly used vegetation indices are the Normalized Difference Vegetation Index (NDVI) and Enhanced Vegetation Index (EVI) [11, 14, 15].

Urban density is another contributing factor to UHIs, estimated using built-up indices. Analyses conducted by [11, 15, 16] discovered a relationship between LST and the Normalized Difference Built-up Index (NDBI).

Most previous studies have relied on regression or correlation analysis to identify associations between contributing factors and UHIs. While correlation analysis is a valid method for estimating the linear relationship between two variables, it cannot establish their causal relationship. Besides, previous studies have often overlooked the temporal relationship between contributing factors and UHIs, which is typically more complex than what can be captured by linear models. Therefore, in this research, we conducted a comprehensive study to investigate whether changes in urban characteristics over time can lead to changes in UHI. To this end, we proposed a statistical test based on the Hotelling T-square test to evaluate the causal relationship between the changes in urban characteristics over time and the changes in UHIs. To this end, we used time series data of precipitation, NDSI, and NDWI to capture the environmental characteristics of urban districts. The NDBI time series was used to model the building density of an urban district, while the amount of green spaces was captured using time series data of NDVI and EVI indices. To assess the intensity of UHI over time, we used the LST time series as an indicator. This research was conducted in Tehran, the capital of Iran. Like other major cities, Tehran faces significant environmental challenges due to the UHI effect. The adverse effects of UHIs have been exacerbated in recent decades due to the increase in the population, urban sprawl, reduction of green space, climate change, non-standard building structures, and air pollution [11, 15, 17]. This research offers valuable insights for Tehran's urban planners and policymakers as they work to mitigate the UHI effect and develop more sustainable and livable urban environments.

The structure of this article is as follows: Section 2 presents the study area, the used data, and the proposed method. In section3, the results of the analysis are presented and discussed. Finally, we draw our conclusion in the fourth section.

## 2. Materials and Methods

### *2.1. Study area*

Tehran Metropolis is located between longitudes 51° to 51° 40' E and latitudes 35° 30' to 35° 51' N, in the northern part of Iran. As the most densely populated city in the country, Tehran covers an area of approximately 700 $km^2$ and is divided into 22 municipal districts [11]. With an approximate population of 11 million, Tehran is the largest city in Iran and ranks 25th among the most populous cities in the world [18]. The Alborz mountain range serves as the northern border of the city, while the Dasht-e-Kavir desert is situated to the south. Therefore, the northernmost areas are at an altitude of about 1800 meters above sea level, while the southernmost areas are at an altitude of approximately 1050 meters, and the central parts of the city are situated at an altitude of around 1200 meters [18]. The climate in Tehran is generally mild in spring and autumn, hot and dry in summer, and cold and relatively wet in winter [11]. The average annual temperature ranges from 15°C to 18°C [11]. The megacity's UHI and air pollution problems are exacerbated by its low level of local wind speed [19, 20]. Figure 1 shows the geographical location of Tehran and its municipal districts.

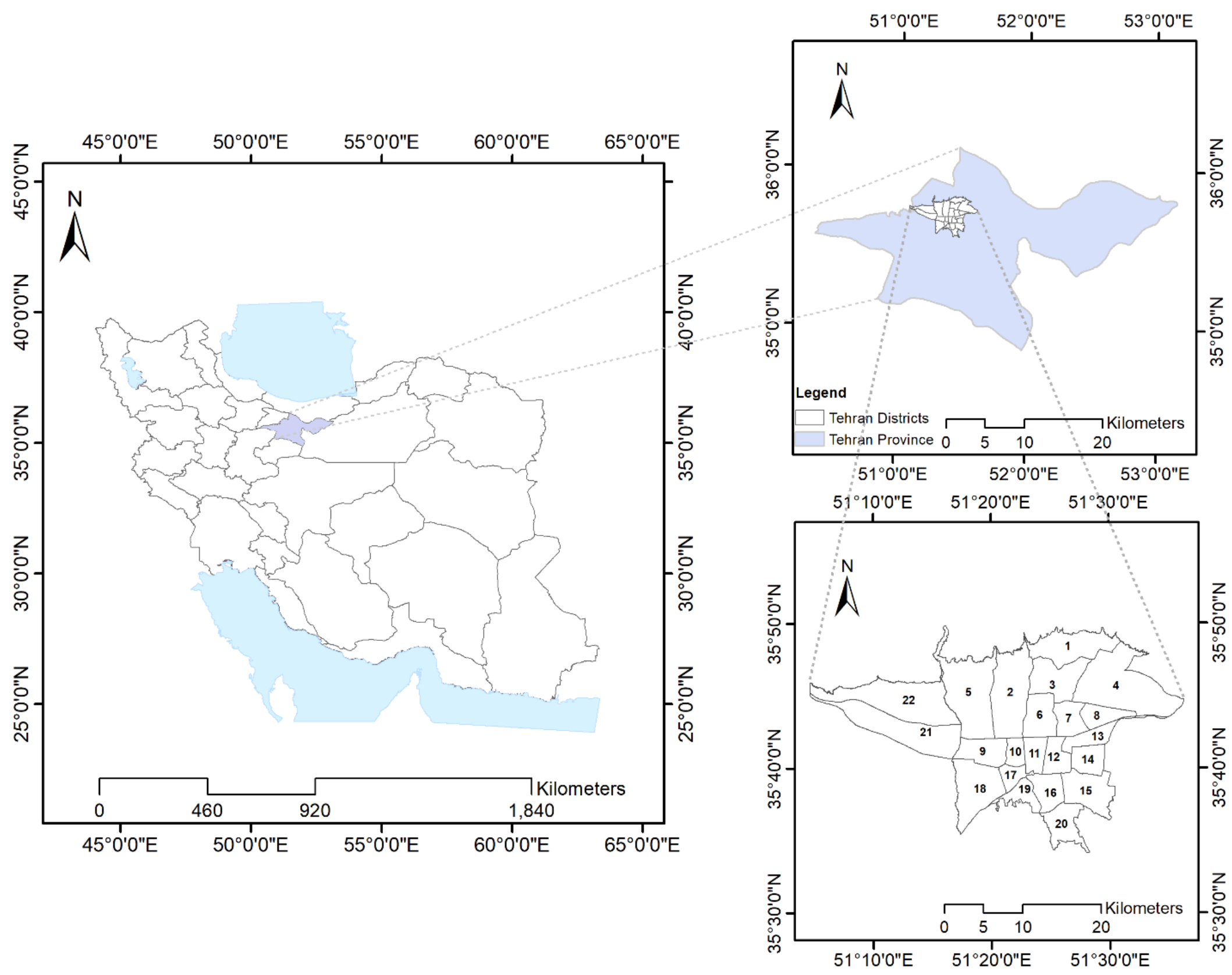

Figure 1- Study area, Tehran districts, Tehran Province, Iran Country

### *2.2. Data*

To better understand the impact of UHI, this research utilizes the Night-LST time series data. This is because, during the day, the Earth is directly exposed to sunlight, which is converted into heat energy [21]. Conversely, at night, there is no sunlight, and any residual energy from the day becomes the sole source of heat [21].

We considered precipitation, NDSI, NDWI, NDBI, EVI, and NDVI as contributing factors to UHI. Precipitation, NDWI, and NDSI capture the environmental and weather characteristics of urban districts. The building density of districts is captured using the NDBI index, while the NDVI and EVI are used to measure green spaces in the city.

We collected precipitation data from CHIRPS with a resolution of 5566 meters, annual Night-LST data, and data for other contributing factors from MODIS-TERRA with a resolution of 1,000 meters. The data was collected annually by averaging daily data for each index within the time range of 2003 to 2021, spanning 19 years. Figure 2 shows the statistical characteristics of the LST time series and all the considered contributing factors.

Data extraction and analysis for this research were performed using Google Earth Engine and Python.

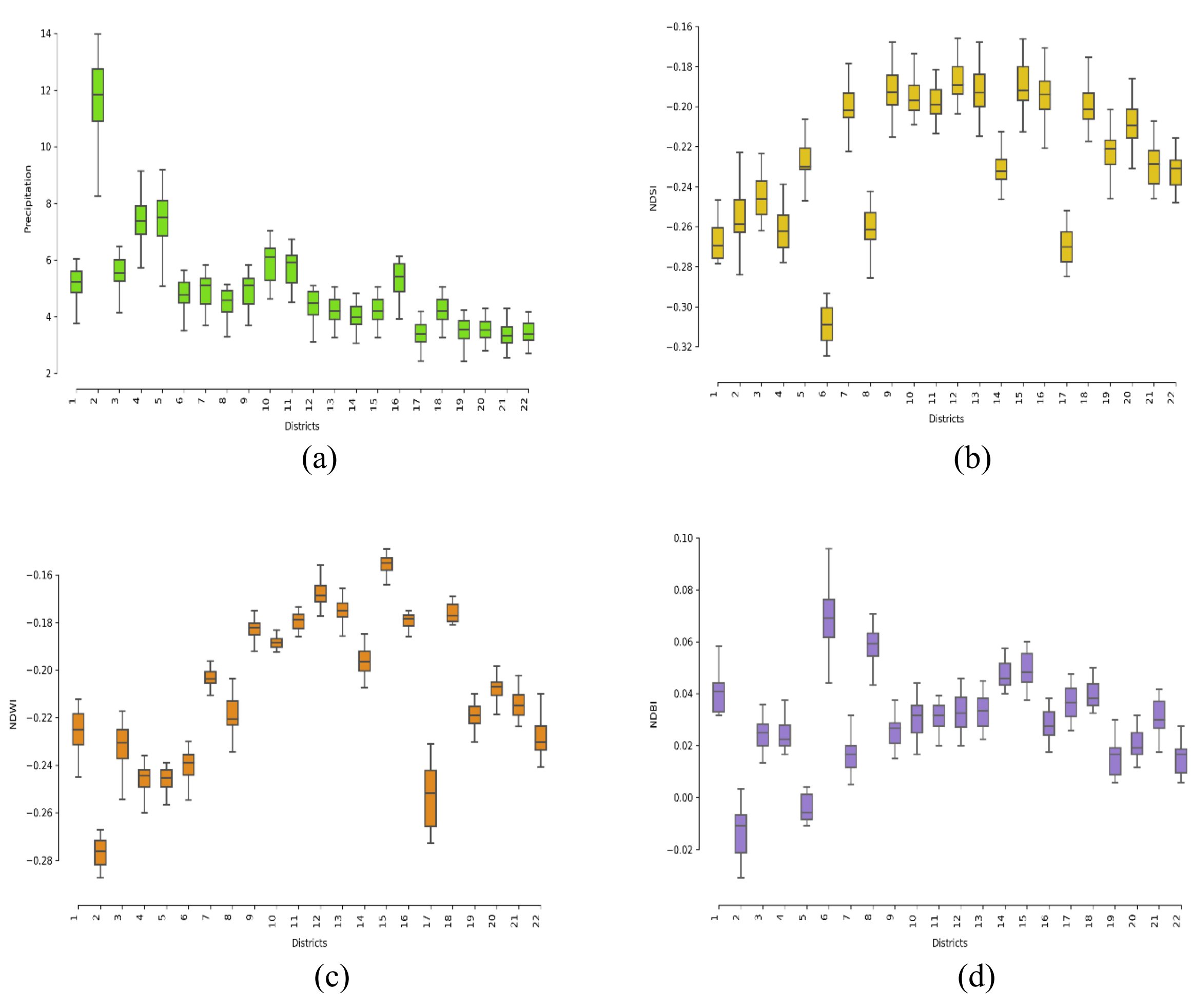

(a) (b)

(c) (d)

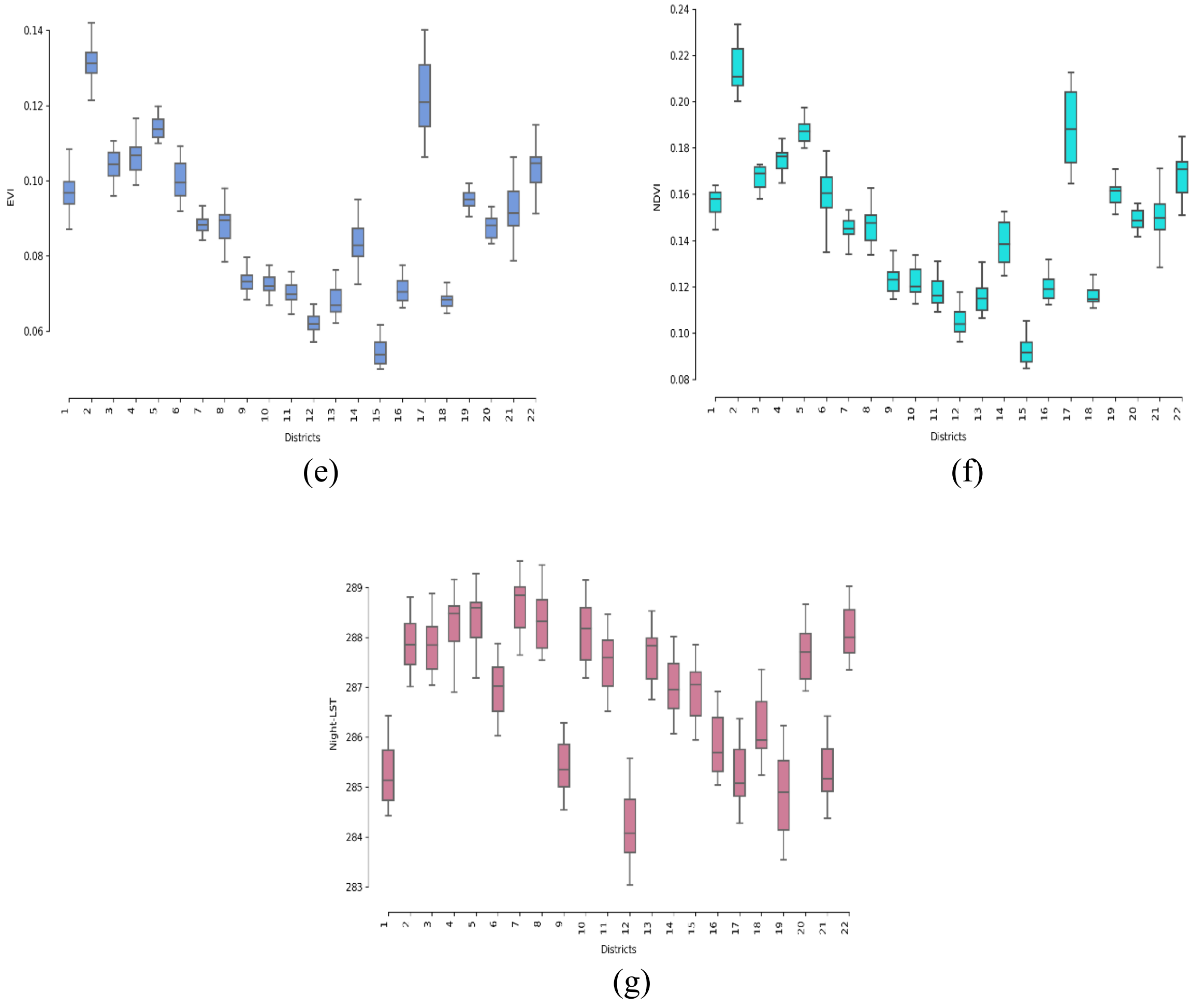

Figure 2- The statistical characteristics of used time series, a) precipitation, b) NDSI, c) NDWI, d) NDBI, e) EVI, f) NDVI, and g) Night-LST

### *2.3. Methods*

We designed two experiments to assess the relationship between UHI and its contributing factors.

In the first experiment, the UHIs of each urban district in Tehran were analyzed. To this end, we first constructed the Night-LST time series by calculating the average Night-LST of the pixels within each urban district for each year in the study period. We used ordinary least squares to estimate the time series trends for each district. Districts were classified into two categories—those with increasing trends and those with non-increasing trends—by comparing their trends to a threshold value. The threshold value for this classification was set to the median of all the estimated trends, which was 0.064. In this experiment, we also evaluated the linear relationship between the Night-LST time series and the time series of each contributing factor by calculating Pearson's correlation coefficient for each district. By employing this approach, we can assess the temporal association between Night-LST and the contributing factors.

In the second experiment, we investigated whether there is a significant causal relationship between the contributing factors and Night-LST across 22 districts over 19 years of study. To perform this analysis, we assumed that if any district characteristics, captured by different indices, contribute to the formation of UHIs during this time, their average values would differ between districts with increasing trends and those with non-increasing trends. To statistically compare the average of the time series for contributing factors between these categories, the Hotelling T-square test was used, which is briefly introduced in the following section.

*2.3.1 Hotelling T- square test*

Hotelling's T-square extends the Student's t-test in the context of multivariate hypothesis testing, allowing for the determination of differences between multivariate averages across different populations [22]. In Hotelling's T-square, the null hypothesis ($H_0$) assumes that there is no significant difference between the average of the groups of districts with a non-increasing trend and those with an increasing trend. On the other hand, the alternative hypothesis ($H_1$) suggests a significant difference between the averages of these groups. In this test, the P-value represents the probability of accepting or rejecting the $H_0$ [23]. A P-value less than or equal to a pre-set threshold (0.01 in our implementation) is considered significant and indicates a causal relationship between the temporal variation of a contributing factor and LST. On the other hand, a P-value greater than this value suggests no significant causal relationship, indicating that the two groups being compared may have similar effects or averages on the LST trend. This test is calculated by Equations (1) and (2) [24]:

$$T^2 = \frac{N_1 N_2}{N_1 + N_2} (\bar{\mathbf{x}}_1 - \bar{\mathbf{x}}_2)^T \mathbf{S^{-1}} (\bar{\mathbf{x}}_1 - \bar{\mathbf{x}}_2) \quad (1)$$

where $\bar{\mathbf{x}}_1$ and $\bar{\mathbf{x}}_2$ represent the sample average of groups one and two, respectively. $\mathbf{S^{-1}}$ is the inverse of the covariance matrix, and, $n = N_1 + N_2 - 2$ where $N_1$ and $N_2$ are sizes of the observations Huang, et al. [25].

$$\frac{n - p + 1}{np} T^2 \sim F_{p,n+1-p} \quad (2)$$

The variable $p$ is dimensional normal distribution, while the $F_{p,n+1-p}$ is an F-distribution [26].

## 3. Results and discussion

### *3.1. Results of the first experiment*

#### *3.1.1 Evaluating the Night-LST Trends*

Figure 3 shows the estimated trends for each district. Based on this Figure, the districts in ascending order of their trends are 5, 7, 22, 13, 10, 2, 3, 20, 6, 8, 11, 15, 9, 16, 14, 17, 1, 21, 18, and 12.

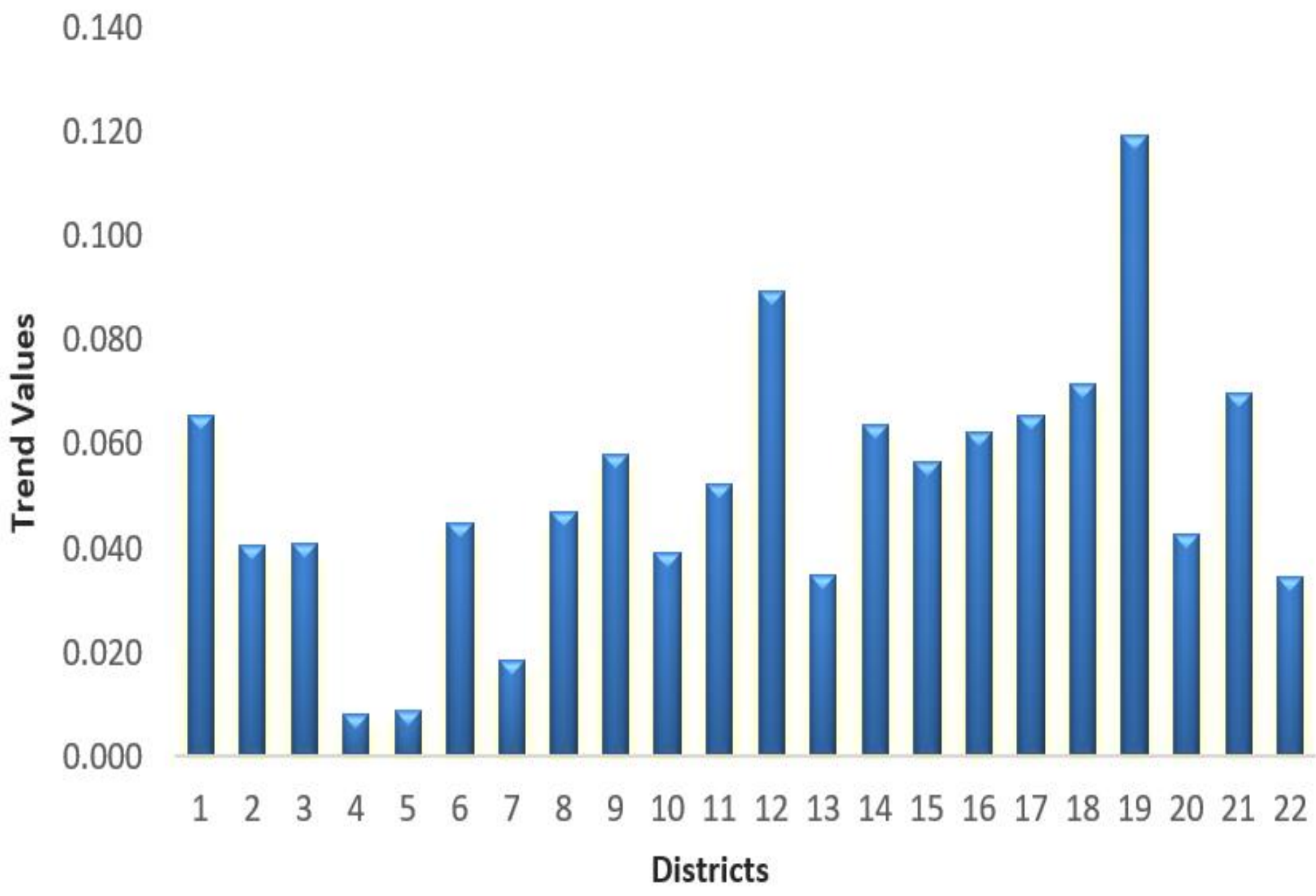

Figure 3- Estimated trends of Night-LST time series for Tehran districts

According to Figure 3, all of the Tehran districts exhibit a positive Night-LST trend, indicating the presence of an increasing UHI effect throughout the city. As observed in these results, District 19, with a trend of 0.119, exhibits the highest increase in Night-LST data and UHI. This is likely because the population of this district is significantly larger relative to its size [27, 28].

On the other hand, District 4 has the lowest trend value among the districts, with a trend of 0.008. This can be attributed to the presence of forest parks and green spaces, as well as its proximity to the Letian Dam, which contributes to the cooler weather in this District.

Furthermore, Districts 2 and 3 exhibit similar Night-LST trends, both having a trend of 0.04. This similarity can be attributed to their adjacent location in northern Tehran, where they experience similar weather conditions. Additionally, some districts exhibit nearly identical trend lines. For instance, Districts 1 and 17 share similar trends, which can be attributed to the presence of abundant green spaces in both Districts. Similarly, Districts 4 and 5 display similar trends due to the presence of forest parks and Azadi Lake. Districts 13 and 22 exhibit similar trends due to their approximately equal population density. Furthermore, Districts 18 and 21, located in the west of Tehran and adjacent to each other, experience similar weather conditions, leading to nearly identical trends.

The classification of trends into two categories showed that Districts 2, 3, 4, 5, 6, 7, 8, 9, 10, 11, 13, 15, 16, 20, and 22 were categorized as the Districts with the non-increasing trends. Meanwhile, Districts 1, 12, 14, 17, 18, 19, and 21 were classified as districts with increasing trends. Figure 4 illustrates the classification results of the districts.

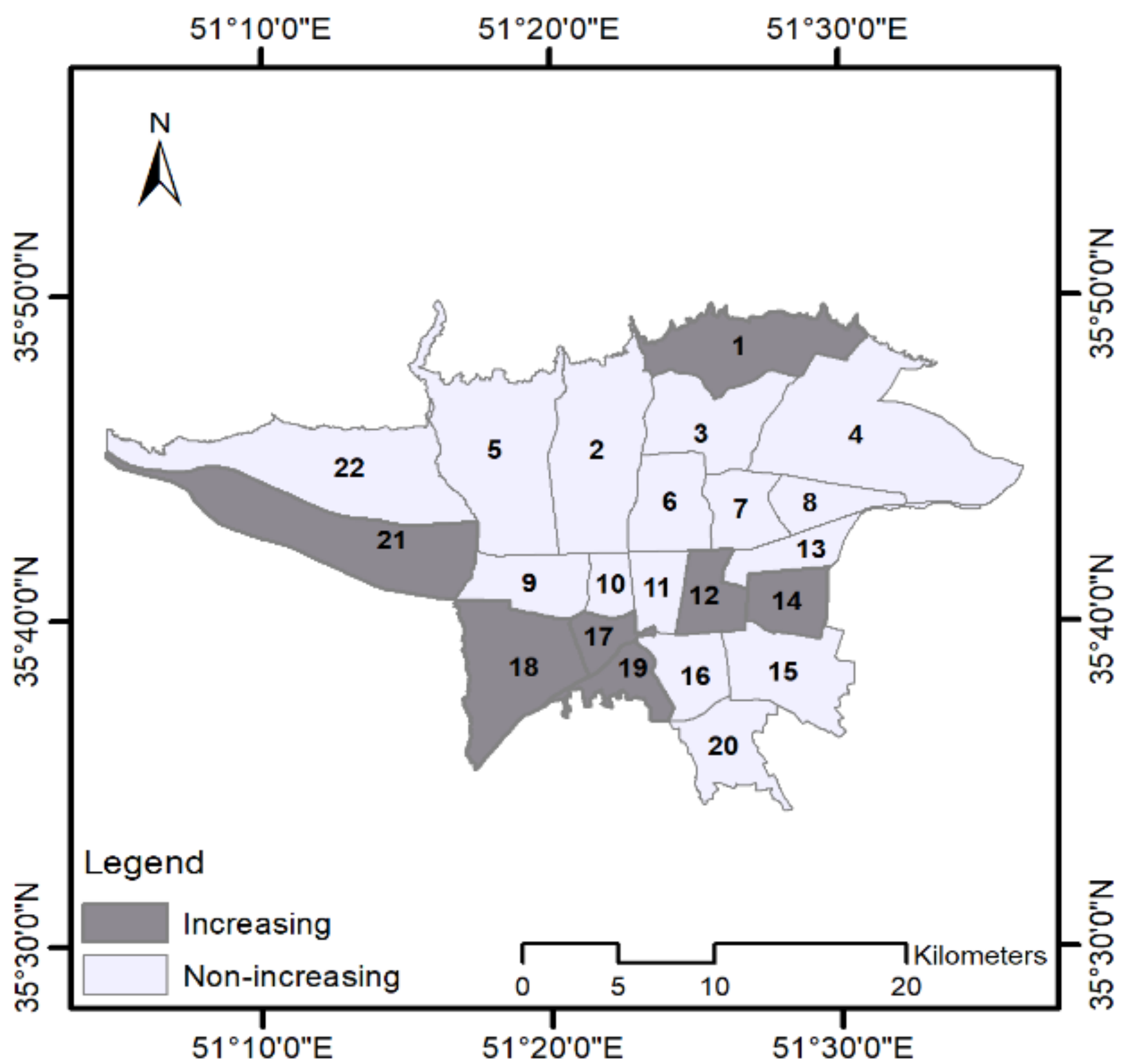

Figure 4 - Classification results of the districts into two categories of increasing and non-increasing trends.

### *3.1.2 Evaluating Pearson's Correlation Coefficient*

The correlation coefficients between the time series of each contributing factor and the Night-LST time series data are listed in Table 1.

Based on Table 1, there is a negative correlation between Night-LST and various factors across multiple districts. Precipitation showed negative correlations with Night-LST for all districts. This is because, despite the increasing trend of UHI, precipitation decreased during the study period. The NDSI also exhibited strong negative correlations for all districts. However, it is noteworthy that cold temperatures in winter can be considered a confounding factor, as they can affect both snow cover and UHI. The NDWI showed a negative correlation in 15 out of 22 districts. This is because these districts mostly contained water bodies, which can help reduce the intensity of UHIs. The EVI and the NDVI exhibit a negative correlation in 12 and 14 out of 22 districts, respectively. This can be interpreted as an increase in urban green spaces in these districts, which helped reduce the UHI effects.

Finally, there is a direct relationship between NDBI and Night-LST. As NDBI values increase, both Night-LST and the UHI also increase. Conversely, a decrease in NDBI values results in a decrease in Night-LST.

Table 1- The correlation coefficients between the time series of contributing factors and Night-LST in each district of Tehran

| Districts | Precipitation | NDSI | NDWI | NDBI | EVI | NDVI |
|---|---|---|---|---|---|---|
| 1 | -0.507 | -0.370 | -0.531 | -0.005 | 0.470 | 0.382 |
| 2 | -0.476 | -0.674 | -0.158 | 0.286 | 0.055 | -0.150 |
| 3 | -0.530 | -0.497 | -0.321 | 0.040 | 0.294 | 0.274 |
| 4 | -0.423 | -0.240 | 0.030 | 0.161 | -0.030 | -0.039 |
| 5 | -0.458 | -0.516 | 0.035 | 0.126 | 0.095 | -0.080 |
| 6 | -0.096 | -0.110 | -0.124 | -0.031 | 0.138 | 0.165 |
| 7 | -0.230 | -0.498 | -0.048 | 0.250 | -0.094 | -0.080 |
| 8 | -0.268 | -0.360 | -0.212 | 0.089 | 0.170 | 0.142 |
| 9 | -0.024 | -0.637 | -0.204 | 0.396 | -0.280 | -0.349 |
| 10 | -0.144 | -0.583 | -0.239 | 0.326 | -0.106 | -0.239 |
| 11 | -0.086 | -0.584 | -0.230 | 0.433 | -0.267 | -0.299 |
| 12 | -0.277 | -0.733 | 0.126 | 0.473 | -0.151 | -0.455 |
| 13 | -0.251 | -0.577 | 0.001 | 0.402 | -0.073 | -0.185 |
| 14 | -0.322 | -0.355 | 0.149 | 0.364 | -0.156 | -0.362 |
| 15 | -0.214 | -0.642 | -0.108 | 0.414 | -0.229 | -0.284 |
| 16 | -0.158 | -0.700 | -0.160 | 0.414 | -0.376 | -0.326 |
| 17 | -0.472 | -0.238 | 0.370 | 0.414 | -0.475 | -0.513 |
| 18 | -0.337 | -0.609 | -0.374 | 0.254 | 0.218 | 0.061 |
| 19 | -0.183 | -0.709 | -0.630 | 0.018 | 0.533 | 0.589 |
| 20 | -0.352 | -0.520 | -0.239 | 0.173 | 0.115 | 0.166 |
| 21 | -0.296 | -0.345 | -0.131 | 0.134 | 0.103 | 0.168 |
| 22 | -0.439 | -0.425 | 0.315 | 0.440 | -0.414 | -0.373 |

Figure 5 illustrates the spatial distribution of the correlation between contributing factors and Night-LST in each district of Tehran. In this Figure, the correlation coefficient for each contributing factor has been divided into five intervals with the natural breaks method.

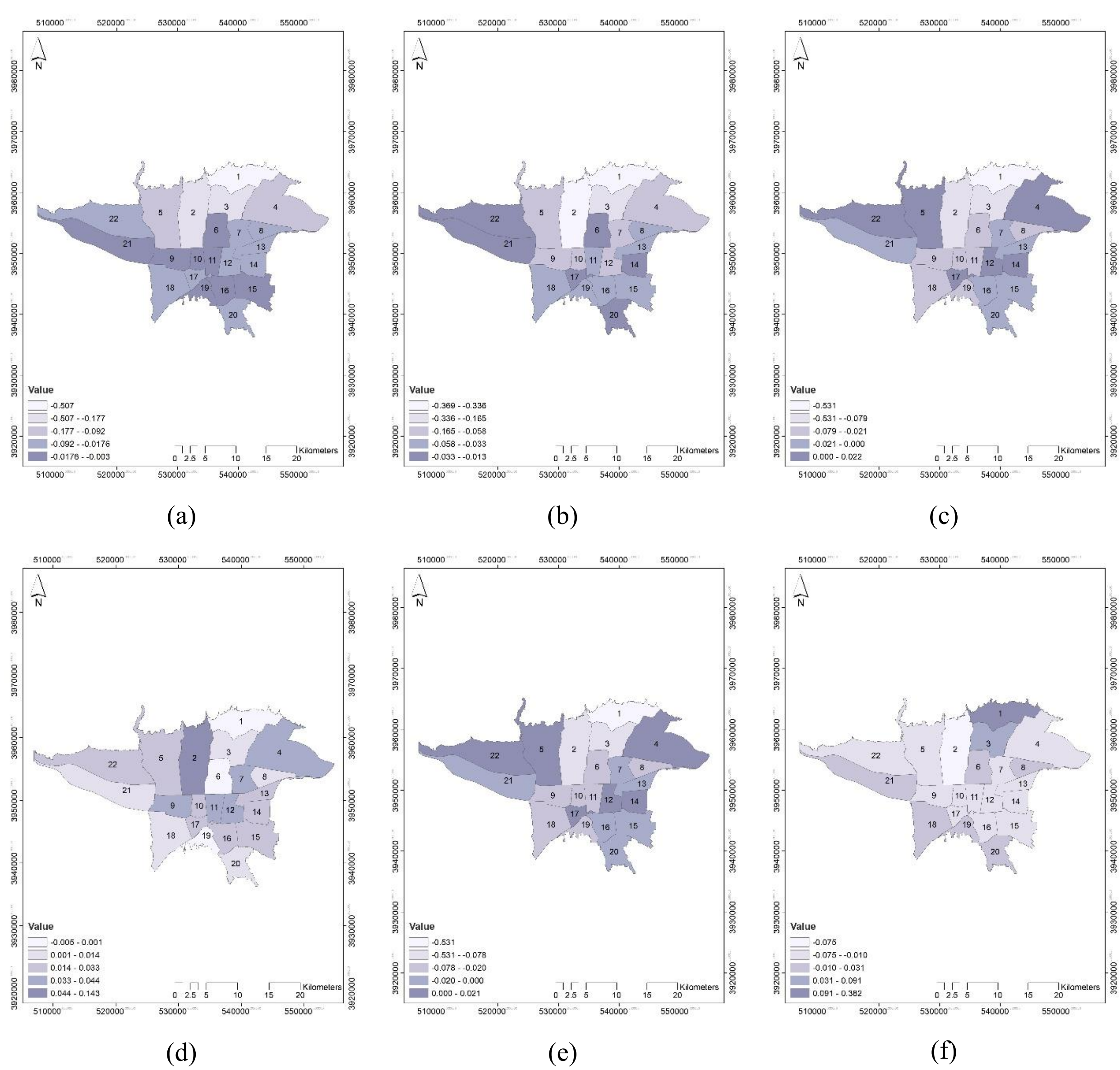

Figure 5- The spatial distribution of correlation of contributing factors in 22 Tehran districts, a) participation, b) NDSI, c) NDWI, d) NDBI, e) EVI, and f) NDVI

### *3.2. Results of the second experiment*

#### *3.2.1 Evaluation of Hotelling T-square*

The results obtained from the Hotteling T-square test are shown in Table 2.

Table 2- The values obtained from the Hotelling T-Square

| Contributing factors | $T^2$ - statistics | P- values |
|---|---|---|
| Precipitation | 122.768 | 0.008 |
| NDSI | 369.822 | 0.003 |
| NDBI | 41.596 | 0.024 |
| NDWI | 43.758 | 0.023 |
| EVI | 1549.052 | 0.001 |
| NDVI | 128.364 | 0.008 |

Based on Table 2, the UHI in Districts with an increasing trend is influenced by several factors compared to the Districts with a non-increasing trend. Precipitation and NDSI have a p-value of less than 0.01, indicating that they can be considered to have a causal association with the UHI. Insufficient precipitation and low NDSI contribute to a rise in LST, thereby amplifying the UHI. Additionally, EVI and NDVI demonstrate a causal role in the increase of UHI, as indicated by their p-values of less than 0.01, with EVI showing the highest T-statistic. Increasing EVI and NDVI during the night can lead to a rise in Night-LST due to the release of carbon dioxide, which further exacerbates the overall UHI intensity. Although the p-values for NDBI and NDWI are greater than 0.01, with NDBI having the highest p-value and the lowest T-statistic, neither NDBI nor NDWI can be considered a significant contributor to the increase in UHI. This is because urbanization has reached its peak, with the primary change being the transformation of smaller, older buildings into larger ones, which has no direct impact on UHI. From the obtained results of NDWI, it can be deduced that the presence and the changes in small-scale artificial water bodies do not have a significant effect on UHI at this scale.

## 4. Conclusions

This research aimed to study the UHIs in 22 Tehran's urban districts between 2003 a 2021 and analyze the roles that different contributing factors play in the formation of UHIs. To this end, we first estimated the trends of the Night-LST time series for each district and calculated the correlation between the Night-LST time series and the time series of each contributing factor. Then we proposed a statistical test based on the Hotteling T-square test to examine the causal relationship between each factor and UHI.

Our findings revealed notable variations in the UHI intensity across districts. Although no district showed a decreasing trend, District 4 had the lowest Night-LST trend, while District 19 had the highest. The estimated correlation revealed a negative association between the temporal variation of the UHI and contributing factors such as precipitation, NDSI, NDWI, EVI, and NDVI. Conversely, a positive correlation was observed between UHI and NDBI. However, correlation analysis can be affected by confounding factors such as winter weather, small water bodies in districts, and subtle changes in urban densities.

This conclusion was supported by the results from the causal analysis, which identified all contributing factors as causal to UHI changes over time, except for NDBI and NDWI.

For future research, exploring non-linear or multi-line trends could enhance the understanding and modeling of UHIs. Additionally, investigating the effectiveness of cooling strategies is recommended. These strategies include modifying building patterns, prioritizing green infrastructure, improving transportation and ventilation systems, and enforcing green construction standards through legislation [11].

**Conflict of interest**
The corresponding author states that there is no conflict of interest.

**Funding**
This research did not receive any specific grant from funding agencies in the public, commercial, or not-for-profit sectors.

**Authors' contributions**
SAS: Data Curation, Writing-Original Draft, visualization, Software
SN: Conceptualization, Methodology, writing- review & editing
AS: writing- review & editing, Supervision,

**References**

[1] L. Howard, *The climate of London: deduced from meteorological observations made in the metropolis and at various places around it*. Harvey and Darton, J. and A. Arch, Longman, Hatchard, S. Highley [and] R. Hunter, 1833.
[2] H. Akbari, "Energy saving potentials and air quality benefits of urban heat IslandMitigation," 2005.
[3] H. Akbari and H. D. Matthews, "Global cooling updates: Reflective roofs and pavements," *Energy and Buildings,* vol. 55, pp. 2-6, 2012.
[4] S. Guhathakurta and P. Gober, "The impact of the Phoenix urban heat island on residential water use," *Journal of the American Planning Association,* vol. 73, no. 3, pp. 317-329, 2007.
[5] A. Almusaed and A. Almusaed, "The urban heat island phenomenon upon urban components," *Biophilic and Bioclimatic Architecture: Analytical Therapy for the Next Generation of Passive Sustainable Architecture,* pp. 139-150, 2011.
[6] A. H. Rosenfeld, H. Akbari, J. J. Romm, and M. Pomerantz, "Cool communities: strategies for heat island mitigation and smog reduction," *Energy and buildings,* vol. 28, no. 1, pp. 51-62, 1998.
[7] T. R. Oke, *Boundary layer climates*. Routledge, 2002.
[8] A. Lilly Rose and M. D. Devadas, "Analysis of land surface temperature and land use/land cover types using remote sensing imagery-a case in Chennai city, India," in *The seventh international conference on urban clim held on*, 2009, vol. 29.
[9] H. Li *et al.*, "A new method to quantify surface urban heat island intensity," *Science of the total environment,* vol. 624, pp. 262-272, 2018.
[10] A. F. Nabizada *et al.*, "Spatial and temporal assessment of remotely sensed land surface temperature variability in Afghanistan during 2000–2021," *Climate,* vol. 10, no. 7, p. 111, 2022.
[11] M. Meftahi, M. Monavari, M. Kheirkhah Zarkesh, A. Vafaeinejad, and A. Jozi, "Achieving sustainable development goals through the study of urban heat island

changes and its effective factors using spatio-temporal techniques: the case study (Tehran city)," in *Natural Resources Forum*, 2022, vol. 46, no. 1: Wiley Online Library, pp. 88-115.
[12] S. Guha, H. Govil, and P. Diwan, "Analytical study of seasonal variability in land surface temperature with normalized difference vegetation index, normalized difference water index, normalized difference built-up index, and normalized multiband drought index," *Journal of Applied Remote Sensing,* vol. 13, no. 2, pp. 024518-024518, 2019.
[13] H. Imran *et al.*, "Impact of land cover changes on land surface temperature and human thermal comfort in Dhaka city of Bangladesh," *Earth Systems and Environment,* vol. 5, pp. 667-693, 2021.
[14] G. Suthar, R. P. Singhal, S. Khandelwal, and N. Kaul, "Spatiotemporal variation of air pollutants and their relationship with land surface temperature in Bengaluru, India," *Remote Sensing Applications: Society and Environment,* vol. 32, p. 101011, 2023.
[15] R. Bala, V. P. Yadav, D. N. Kumar, and R. Prasad, "Exploring the relationship of land surface parameters and air pollutants with land surface temperature in different cities using satellite data," *Advances in Space Research,* 2024.
[16] N. Rashid, J. M. Alam, M. A. Chowdhury, and S. L. U. Islam, "Impact of landuse change and urbanization on urban heat island effect in Narayanganj city, Bangladesh: A remote sensing-based estimation," *Environmental Challenges,* vol. 8, p. 100571, 2022.
[17] H. Rezaeei Rad and M. Rafieyan, "Estimating the spatial-temporal Changes in intensity of the heat island in Tehran Metropolitan by Using ASTER and Landsat8 Satellite Images," *Regional Planning,* vol. 7, no. 27, pp. 47-60, 2017.
[18] S. Nasehi, A. Yavari, and E. Salehi, "Investigating the spatial distribution of land surface temperature as related to air pollution level in Tehran metropolis," *Pollution,* vol. 9, no. 1, pp. 1-14, 2023.
[19] H. Malakooti and A. Bidokhti, "A modeling study of boundary layer wind flow over Tehran region during a high pollution episode," *Journal of Applied Fluid Mechanics,* vol. 7, no. 2, pp. 299-313, 2014.
[20] A. Bidokhti, Z. Shariepour, and S. Sehatkashani, "Some resilient aspects of urban areas to air pollution and climate change, case study: Tehran, Iran," *Scientia Iranica. Transaction A, Civil Engineering,* vol. 23, no. 5, p. 1994, 2016.
[21] S. M. Mirou, W. Zeiada, R. I. Al-Ruzouq, and R. N. Hassan, "Investigation of Diurnal and Seasonal Land Surface Temperature," in *2022 Advances in Science and Engineering Technology International Conferences (ASET)*, 2022: IEEE, pp. 1-6.
[22] R. M. Ueda and A. M. Souza, "An effective approach to detect the source (s) of out-of-control signals in productive processes by vector error correction (VEC) residual and Hotelling's T2 decomposition techniques," *Expert Systems with Applications,* vol. 187, p. 115979, 2022.
[23] M. S. Thiese, B. Ronna, and U. Ott, "P value interpretations and considerations,"

*Journal of thoracic disease,* vol. 8, no. 9, p. E928, 2016.
[24] D. C. Montgomery, *Introduction to statistical quality control*. John Wiley & Sons, 2020.
[25] Y. Huang, C. Li, R. Li, and S. Yang, "An overview of tests on high-dimensional means," *Journal of Multivariate Analysis,* vol. 188, p. 104813, 2022.
[26] M. Chandrajit, R. Girisha, and T. Vasudev, "Motion Segmentation from Surveillance Video using modified Hotelling's T-Square Statistics," *International Journal of Image, Graphics and Signal Processing,* vol. 8, no. 7, p. 41, 2016.
[27] H. Sharifi, M. R. Masjedi, H. Emami, M. Ghanei, and S. Buist, "Burden of obstructive lung disease study in tehran: research design and lung spirometry protocol," *International Journal of Preventive Medicine,* vol. 5, no. 11, p. 1439, 2014.
[28] S. Najjarzadeh and M. Zaeimdar, "Investigating the relationship between urban green spaces and population in the 22 districts of Tehran city.," *Environmental Management and Sustainable Development,* vol. 1, pp. 0-0, 2018. [Online]. Available: https://sid.ir/paper/513592/fa.